\documentclass[openany, twocolumn]{aastex7}

\usepackage{amssymb}
\usepackage{amsmath}
\usepackage[frozencache,cachedir=.]{minted}

\definecolor{LightGray}{gray}{0.9}

\newcommand{\secref}[1]{Section~\ref{#1}}

\newcommand{\inlinepy}[1]{\mintinline{python}{#1}}

\begin{document}

\title{\textsc{dolphin}: A fully automated forward modeling pipeline powered by artificial intelligence for galaxy-scale strong lenses}

\author[orcid=0000-0002-5558-888X,sname='Shajib']{Anowar J.~Shajib}
\altaffiliation{NHFP Einstein Fellow}
\affiliation{Department of Astronomy \& Astrophysics, University of Chicago, Chicago, IL 60637, USA}
\affiliation{Kavli Institute for Cosmological Physics, University of Chicago, Chicago, IL 60637, USA}
\affiliation{Department of Physics \& Astronomy, University of California, Los Angeles, CA 90095, USA}
\affiliation{Center for Astronomy, Space Science and Astrophysics, Independent University, Bangladesh, Dhaka 1229, Bangladesh}
\email[show]{ajshajib@uchicago.edu}

\author[sname=Nihal]{Nafis Sadik Nihal} 
\affiliation{Center for Astronomy, Space Science and Astrophysics, Independent University, Bangladesh, Dhaka 1229, Bangladesh}
\email{nafissadik7@iut-dhaka.edu}

\author[sname=Tan]{Chin Yi Tan}
\affiliation{Department of Physics, University of Chicago, Chicago, IL 60637, USA}
\affiliation{Kavli Institute for Cosmological Physics, University of Chicago, Chicago, IL 60637, USA}
\email{chinyi@uchicago.edu}

\author{Vedant Sahu}
\affiliation{Department of Physics \& Astronomy, University of California, Los Angeles, CA 90095, USA}
\email{vedantsahu@gmail.com}

\author{Simon Birrer}
\affiliation{Department of Physics and Astronomy, Stony Brook University, Stony Brook, NY 11794, USA}
\email{simon.birrer@stonybrook.edu}

\author{Tommaso Treu}
\affiliation{Department of Physics \& Astronomy, University of California, Los Angeles, CA 90095, USA}
\email{tt@astro.ucla.edu}

\author{Joshua Frieman}
\affiliation{Department of Astronomy \& Astrophysics, University of Chicago, Chicago, IL 60637, USA}
\affiliation{Kavli Institute for Cosmological Physics, University of Chicago, Chicago, IL 60637, USA}
\email{jfrieman@uchicago.edu}

\begin{abstract}

Strong gravitational lensing is a powerful tool for probing the internal structure and evolution of galaxies, the nature of dark matter, and the expansion history of the Universe, among many other scientific applications. For almost all of these science cases, modeling the lensing mass distribution is essential. For that, forward modeling of imaging data to the pixel level is the standard method used for galaxy-scale lenses. However, the traditional workflow of forward lens modeling necessitates a significant amount of human investigator time, requiring iterative tweaking and tuning of the model settings through trial and error. An automated lens modeling pipeline can substantially reduce the need for human investigator time. In this paper, we present \textsc{dolphin}, an automated lens modeling pipeline that combines artificial intelligence with the traditional forward modeling framework to enable full automation of the modeling workflow. \textsc{dolphin} uses a neural network model to perform visual recognition of the strong lens components, then autonomously sets up a lens model with appropriate complexity, and fits the model with the modeling engine, \textsc{lenstronomy}. Thanks to the versatility of \textsc{lenstronomy}, \textsc{dolphin} can autonomously model both galaxy--galaxy and galaxy--quasar strong lenses. 

\end{abstract}



\section{Introduction}

Strong lensing occurs when a massive object, such as a galaxy or galaxy cluster, lies between a distant light source and an observer. The gravitational field of the massive object bends the light from the source, creating multiple images of the source for the observer. The positions and shapes of these images depend on the mass distribution of the lensing object and the geometric configuration of the lens, source, and observer. Consequently, strong lensing is a powerful tool for probing several key questions in cosmology and astrophysics. For example, it provides a unique opportunity to study the internal structure and evolution of galaxies at intermediate redshifts ($z \sim 0.5$) \citep{Shajib24}, the nature of dark matter \citep{Vegetti24}, and the expansion history of the Universe \citep{Birrer24}.

Imaging data, particularly in infrared to ultraviolet wavelengths, provides the most accessible type of strong lensing observables. Traditionally, forward modeling of this imaging data to constrain the mass distribution of the lensing object is an essential step for almost all the science applications of strong lensing. At the galaxy scale, it is a standard procedure to model the imaging data up to the pixel level. In contrast, at the cluster scale, the lensing observables are summarized into astrometric positions of the lensed galaxies or knots for constraining the lensing mass distribution. Several forward modeling software programs are used in the literature, for example, \textsc{lenstronomy} \citep{Birrer18, Birrer21b}, \textsc{pyAutoLens} \citep{Nightingale18}, \textsc{Glee} \citep{Suyu10}, \textsc{glafic} \citep{Oguri10b}, \textsc{lenstool} \citep{Kneib11}. On top of being computationally intensive, forward modeling of strong lensing imaging data is also a time-consuming process requiring significant human intervention. The process involves several steps, such as visual identification of the lens and source components, selection of appropriate model components for them, and the evaluation of goodness-of-fit, which may lead to further fine-tuning of the model setup, with all these steps potentially repeated for many times until a sufficiently good lens model is achieved \citep[e.g.,][]{Shajib19}.

The significant requirement of investigator time for lens modeling has motivated the development of automated techniques that can perform the modeling process more efficiently. Especially, the ongoing and upcoming large-area sky surveys, such as \textit{Euclid}, the Rubin Observatory Legacy Survey of Space and Time (LSST), and the \textit{Roman} Space Telescope, will increase the sample sizes of known lenses by two orders of magnitudes or more \citep{Oguri10, Collett15, Abe24}, making automated lens modeling algorithms essential. For example, a decision-tree-based algorithm \citep{Shajib19} can be used to automate the fine-tuning of the model setup (e.g., \citealt{Schmidt22}, using \textsc{lenstronomy}; \citealt{Ertl23}, using \textsc{Glee}; automated selection of a subset of the model settings and the subsequent optimization is also performed by \textsc{pyAutoLens}). However, these automated algorithms still require a set of initial inputs from a human investigator to initiate the lens model before successfully obtaining a good lens model. Recently, machine learning (ML) algorithms have also been used for the extraction of the lens model parameters, or parameters that directly relate to galaxy properties or cosmology \citep[e.g.,][]{Hezaveh17, Poh22, Erickson24, Euclid25}. These ML algorithms have built-in automation, given that, once properly trained, the neural networks (NNs) can extract the parameters without further human intervention in the extraction process. However, for accurate and precise parameter extraction, a significant amount of effort and time investment from a human investigator may not be avoided when preparing the training data set.

In this work, we present \textsc{dolphin}\footnote{\url{https://github.com/ajshajib/dolphin}}, an automated lens modeling pipeline that integrates artificial intelligence (AI) with forward modeling, offering a hybrid approach for fully automating the lens modeling process. \textsc{dolphin} uses an NN to perform visual recognition of the strong lens components to inform setting up an optimal lens model, which is then fitted with a conventional forward modeling software program. \textsc{dolphin} uses \textsc{lenstronomy} as its modeling engine, which is a versatile and efficient software package for modeling strong lenses. Furthermore, \textsc{lenstronomy}\footnote{\url{https://github.com/lenstronomy/lenstronomy}} has the unique combination of being open source and publicly available, and it is capable of modeling both galaxy--quasar and galaxy--galaxy lens systems. Thus, \textsc{dolphin}, while also being open source, leverages the unique features of \textsc{lenstronomy} to provide an automated modeling pipeline for both galaxy--quasar and galaxy--galaxy lenses. \textsc{dolphin} has been first developed and applied on samples of the galaxy-galaxy lenses in a semi-automated mode, where the visual recognition was performed by human modelers to set up the lens models \citep{Shajib21, Tan24}. In this paper, we complete the AI module of \textsc{dolphin} to transform it into a \textit{fully} automated modeling pipeline.

\textit{Our presented methodology to fully automate lens modeling by combining AI with a forward modeling framework is the first in the literature.} This hybrid approach has several advantages over purely ML-based approaches. First, our approach still uses the conventional forward modeling approach for the final optimization of the lens model, which is a well-established and reliable method for lens modeling. This approach is based on the full likelihood of the data, which leads to an optimal extraction of the information. Second, although the NN approach will be paramount for sample sizes of $\gtrsim\mathcal{O}(10^5)$, our automated hybrid approach will provide a way to validate these neural-network-based models for a smaller subset with the traditional forward modeling approach.

The paper is organized as follows. In \secref{sec:overall_workflow}, we provide an overview of \textsc{dolphin}'s workflow and associated modules, and demonstrate its automated modeling capability with an example lens system. In \secref{sec:ml}, we describe \textsc{dolphin}'s AI that is trained to perform visual recognition of lens components in the imaging data. We conclude the paper with a discussion in \secref{sec:conclusion}.

\begin{figure*}
    \centering
    \includegraphics[width=\textwidth]{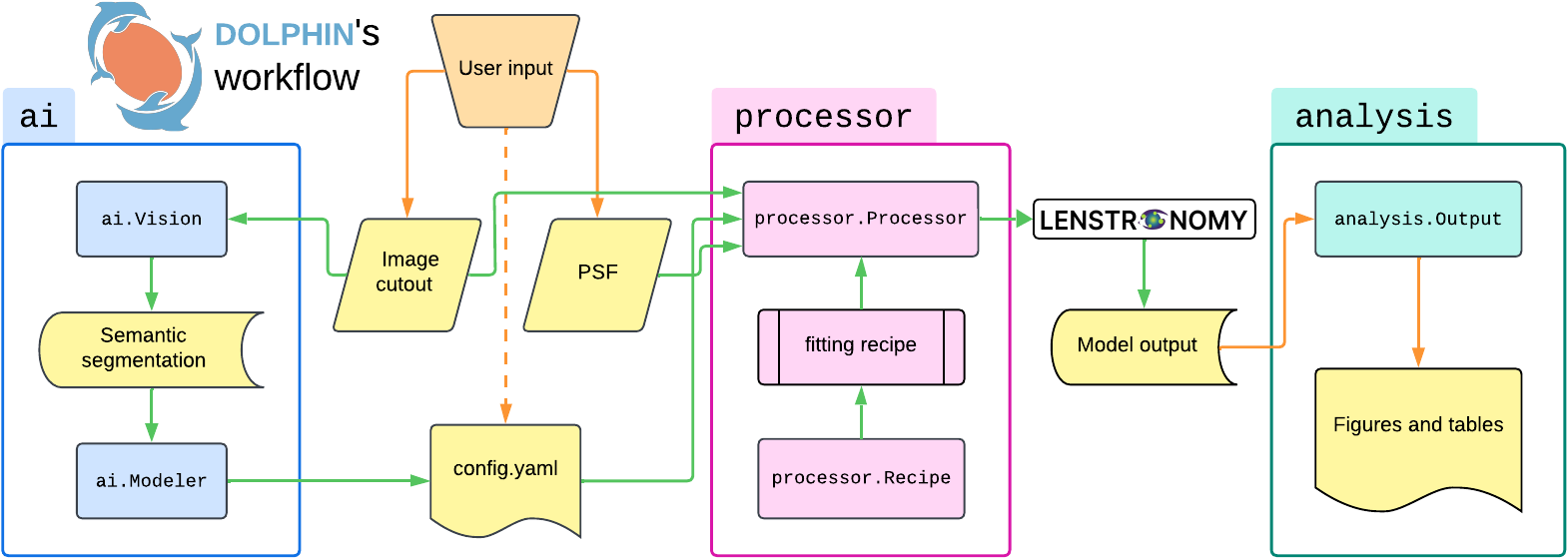}
    \caption{\label{fig:dolphin_workflow}
        The overall workflow of \textsc{dolphin}. The green arrows represent automated procedures performed by \textsc{dolphin}, and the orange arrows represent manual procedures performed by a user. The yellow shapes represent products that are stored on the hard drive. The workflow consists of three main modules: \inlinepy{ai}, \inlinepy{processor}, and \inlinepy{analysis}. The \inlinepy{ai.Vision} module contains a trained NN model that identifies the lens components in the imaging data (i.e., semantic segmentation). The \inlinepy{ai.Modeler} then sets up the lens model and optimization configurations in the \texttt{config.yaml} file. This \texttt{config.yaml} file can be further tweaked manually by the user (depicted with the dashed orange arrow), if necessary. The \inlinepy{processor} module then optimizes the lens model with the \textsc{lenstronomy} modeling engine employing the fitting recipe provided by \texttt{processor.Recipe} class, and then saves the model output. Finally, the \inlinepy{analysis} module provides tools to the user to analyze the model outputs and produce the desired figures and tables.}
\end{figure*}

\section{Overview of \textsc{dolphin}'s workflow} \label{sec:overall_workflow}

In this section, we briefly describe the workflow of \textsc{dolphin} and the associated modules. The overview of \textsc{dolphin}'s workflow is illustrated in Fig. \ref{fig:dolphin_workflow}. The pipeline is designed to be fully automated, where the user is only required to provide the imaging data and the initial or accurate point spread function (PSF) for the lens systems to be modeled. We give an overview of the main modules in Section \ref{sec:main_modules}, describe the fitting recipes in Section \ref{sec:fitting_recipe}, and provide an example of lens modeling with \textsc{dolphin} in Section \ref{sec:running_pipeline}.

\subsection{Description of the main modules} \label{sec:main_modules}

In this section, we provide a general overview of the workflow of \textit{dolphin}'s automated modeling procedure. \textsc{dolphin} consists of three main modules as illustrated in Fig. \ref{fig:dolphin_workflow}:

\begin{enumerate}
    \item \inlinepy{ai}: This module contains a NN model for semantic segmentation, that is, identifying the lens components in the imaging data and classifying them. The segmentation is then processed by the \inlinepy{ai.Modeler} class to set up an optimal lens model. This setting up of an optimal model is largely based on the decision tree algorithm developed by \citet{Shajib19}. In essence, \inlinepy{ai.Modeler} extracts from the segmentation the numbers and positions of the lens components, e.g., the central nd satellite deflectors, the quasar images, then sets up the model incorporating these components, and finally saves the model configuration in a \texttt{config.yaml} file.
    \item \inlinepy{processor}: This module takes the provided model setup and the class \texttt{processor.Config} sets up the lens model in the specifications required by \textsc{lenstronomy}. Then, this module optimizes the model using optimizers such as Particle Swarm Optimization \citep[PSO;][]{Kennedy95} or Markov Chain Monte Carlo (MCMC). The PSO optimization routine is crafted in the \inlinepy{processor.Recipe} class, which guides the initial optimization of the lens model for a more efficient convergence towards a stable solution (described in Section \ref{sec:fitting_recipe}). The class \inlinepy{processor.Processor} then runs the fitting recipe for the given model configuration and saves the model output. 
    \item \inlinepy{analysis}: This module (i.e., the class \inlinepy{analysis.Output}) provides some quality-of-life tools to analyze the results of the lens modeling, such as the goodness-of-fit, the convergence of the optimizer, and the distribution of the lens model parameters. 
\end{enumerate}

\subsection{Fitting recipes} \label{sec:fitting_recipe}

Although a single PSO optimization can, in principle, provide a solution after a sufficient number of iterations, the convergence speed of the optimizer can be improved by using a fitting recipe that guides the optimization process. For example, instead of all the model parameters being free at the beginning, the optimization can be iteratively guided toward convergence by allowing subsets of the parameters to be free while keeping the others fixed at assumed or previously optimized values (from a previous iteration). The fitting recipe is a set of instructions that guides the optimization process to a stable solution. An iterative fitting recipe is also necessary when the PSF is required to be iteratively constructed as done for galaxy--quasar systems \citep[e.g.,][]{Shajib19, Schmidt22}. In addition to the \texttt{"galaxy-quasar"} fitting recipe, a purpose-built \texttt{"galaxy-galaxy"} fitting recipe is also provided for galaxy--galaxy lensing systems \citep{Shajib21, Tan24}.


 \subsection{Running the pipeline} \label{sec:running_pipeline}

The user first needs to prepare the image cutouts and the PSFs for each lens system in \texttt{hdf5} files that include the specific information and datasets required by \textsc{lenstronomy}. These specifications of the input files, along with the desired directory structure (i.e., \texttt{io\_directory}) to store the input and output files, are described in \textsc{dolphin}'s documentation. The lens model specifications of each lens are described in a \texttt{yaml} file (i.e., \texttt{config.yaml}) file stored within the \texttt{io\_directory}. In the fully automated pipeline, these \texttt{yaml} files are generated by the AI. However, a user can produce these \texttt{yaml} files to run \textsc{dolphin} in a semi-automated mode, or tweak the AI-generated \texttt{yaml} files as necessary before running the model optimization.

 As an example, the following \textsc{python} code runs \textsc{dolphin} for the quadruply lensed quasar system J2205$-$3727 \citep{Lemon23}:

\begin{minted}[
    breaklines,
    frame=lines,
    framesep=2mm,
    baselinestretch=1.2,
    bgcolor=LightGray,
    fontsize=\footnotesize,
    % linenos
    ]{python}
from dolphin.ai import Vision
from dolphin.ai import Modeler
from dolphin.processor import Processor

io_directory_path = "path/to/io_directory"

vision = Vision(io_directory_path, source_type="quasar")
vision.create_segmentation_for_single_lens( lens_name="J2205-3727", band_name="F814W")

modeler = Modeler(io_directory_path)
modeler.create_config_for_single_lens( lens_name="J2205-3727", band_name="F814W")

processor = Processor(io_directory_path)
processor.swim(lens_name="J2205-3727", model_id="example", recipe_name="galaxy-quasar")
\end{minted}

\textit{The few lines of code above are all a user needs to run an automated lens model with \textsc{dolphin}.} This makes for a huge contrast with the amount of code needed to set up and run a lens model manually, for example, with \textsc{lenstronomy}\footnote{\href{https://github.com/TDCOSMO/WGD2038-4008/blob/da1eda8a0c5c111cc4a59ca9d5e94f31bf3cac02/lenstronomy_modeling/notebooks/Multiband\%20Image\%20Modeling.ipynb}{A \textsc{Jupyter} notebook with manual lens modeling, for example}}. 
After running the pipeline, the user can plot an overview of the lens model by running the following code:

\begin{minted}[
    breaklines,
    frame=lines,
    framesep=2mm,
    baselinestretch=1.2,
    bgcolor=LightGray,
    fontsize=\footnotesize,
    % linenos
    ]{python}
from dolphin.analysis import Output

output = Output(io_directory_path)
fig = output.plot_model_overview(lens_name="J2205-3727", model_id="example")
\end{minted}

\begin{figure*}
    \includegraphics[width=\textwidth]{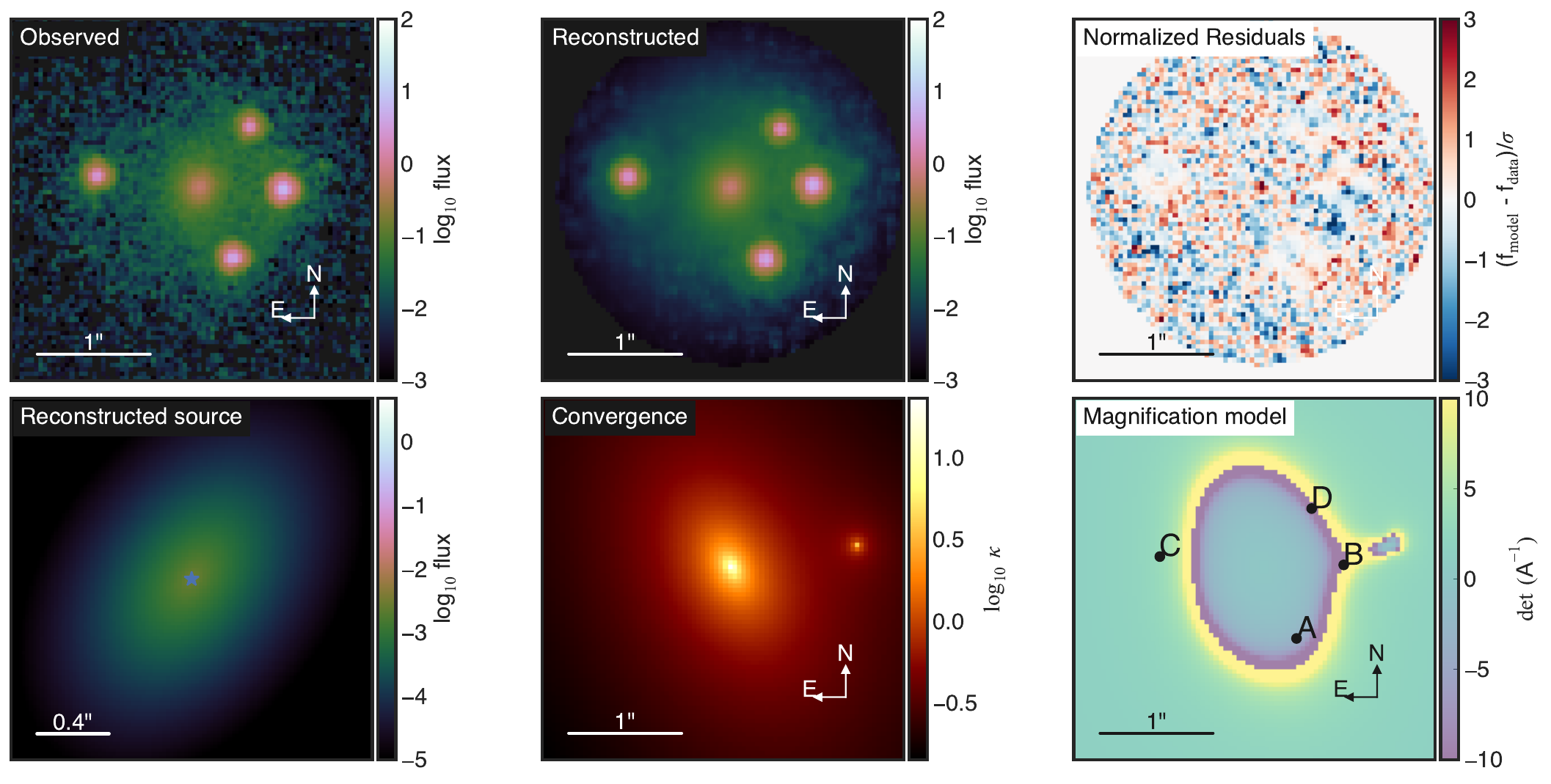}
    \caption{\label{fig:model_overview}
        An overview of the lens model for the galaxy--quasar system J2205$-$3727. The top row shows the observed image cutout in the F814W filter, the optimized-model-based reconstruction, and the residual image. The bottom row shows the reconstructed flux distribution of the host galaxy, the convergence, and the magnification model. The model is optimized with PSO, and the PSF is iteratively constructed with the \texttt{"galaxy-quasar"} fitting recipe from \texttt{processor.Recipe} class.
        }
\end{figure*}
For the modeling here, we use the image cutout and initial PSF estimate for the system J2205$-$3727 from \citet{Shajib19, Schmidt22}. Figure \ref{fig:model_overview} shows the overview of the optimized lens model that is fitted to the noise level, as produced by the \inlinepy{output.plot_model_overview()} function. Notably, \textsc{dolphin}'s AI (described next in Section \ref{sec:ml}) has correctly identified a satellite deflector slightly northwest of the westernmost quasar image and has included it in the lens model, as can be noticed in the `Convergence' panel of Figure \ref{fig:model_overview}. Here, \textsc{dolphin} models the mass distribution of the central deflector with an elliptical power law \citep[EPL;][]{Tessore15}, and that of the satellite deflector with a singular isothermal ellipsoid \citep[SIE;][]{Kormann94}. Additionally, a residual shear field \citep[or, external shear;][]{Shajib24} is added to the mass model. The light profiles of the central and satellite deflectors are modeled with S\'ersic functions \citep{Sersic68}. The model is optimized with PSO for 100 iterations with 100 particles. We use the \texttt{"galaxy-quasar"} fitting recipe from the \inlinepy{processor.Recipe}, which includes the PSF reconstruction step in the optimization process \citep[e.g., as performed in][]{Shajib19}. We do not run the MCMC sampling here, as the PSO optimization is sufficient to illustrate that \textsc{dolphin} can autonomously set up the lens model and optimize the model to produce a stable solution. In real use cases of \textsc{dolphin}, a user may run the MCMC sampling to obtain the posterior distribution of the lens model parameters.

\section{ML-based visual recognition of lens components} \label{sec:ml}

In this section, we describe the ML-based visual recognition model that identifies the lens components in the imaging data within the \inlinepy{ai.Vision} class. We describe the network architecture and activation functions in \secref{sec:network}, the training dataset in \secref{sec:training_set}, the training procedure in \secref{sec:training}, and the validations and tests in \secref{sec:validation}.


\subsection{NN architecture} \label{sec:network}

We implement a U-Net architecture using \textsc{tensorflow} and \textsc{Keras} \citep{Chollet15, Abadi16}. The U-Net architecture is a popular choice for semantic segmentation, which consists of an encoder path (down-sampling) and a decoder path (up-sampling) \citep{Ronneberger15}. 

\begin{figure*}
    \includegraphics[width=\textwidth]{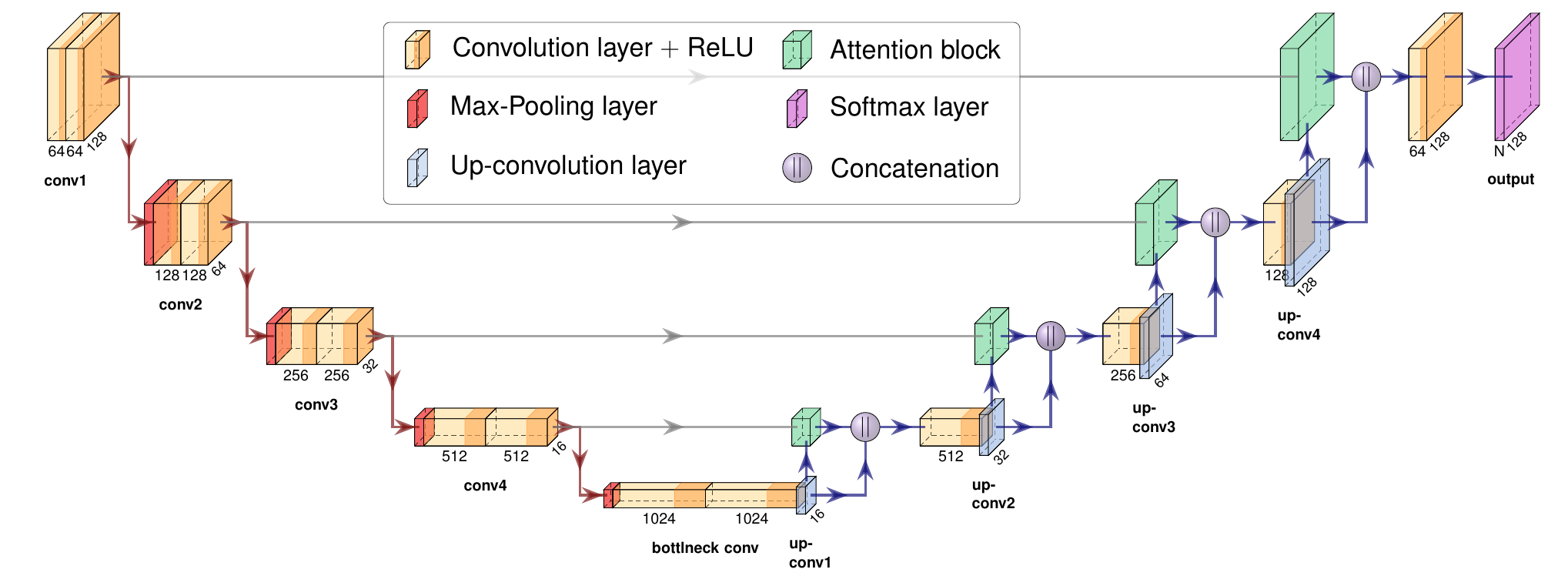}
    \caption{\label{fig:unet_architecture}
        U-Net architecture of our NN to perform semantic segmentation of imaging data for galaxy-scale strong lenses. The U-Net model consists of an encoder path (down-sampling, red arrows) and a decoder path (up-sampling, blue arrows). The encoder path consists of convolutional layers followed by max pooling layers. We use two consecutive convolutional layers in each block with a dropout layer in between, which is not illustrated in this diagram. The decoder path consists of up-sampling layers followed by convolutional layers. The output layer has $N=4$ (for galaxy--galaxy lenses) or $N=5$ (for galaxy--quasar lenses) channels, each providing the probability of one of these classes for every pixel: background, central deflector, quasar image, the lensed arc, and satellite of the central deflector.}
    \end{figure*}

Our model (shown in Figure  \ref{fig:unet_architecture}) takes single-band images of size 128$\times$128 as inputs. Future upgrades of \textsc{dolphin} will include new NN models to allow multi-band imaging for improved performance in segmentation. The encoder path in the U-Net employs two $3\times3$ convolutional layers followed by a $2\times2$ Max-Pooling layer in each downsampling step. Each upsampling step in the decoder path consists of a $2\times2$ transposed convolution, an attention block \citep{Oktay2018}, and then a $3\times3$ convolutional layer. The attention block is applied to the feature maps from both the encoder and decoder paths, and its output is concatenated with the feature map from the decoder path before passing on to the next layer. Each convolutional layer has the rectified linear unit (ReLU) activation function to induce non-linearity \citep{Nair10}. To prevent overfitting, we add dropout layers between the two convolutional layers in each block of the encoder path, with a rate of 0.1 for the first two layers, 0.2 for the next two layers, and increasing to 0.3 for the bottleneck layer.


The attention block we use to filter relevant features is a soft attention mechanism. The encoder and decoder features are transformed using 1$\times$1 convolutions. Then, we add the transformed features and apply the ReLU activation function. Next, an attention mask is generated using a sigmoid activation function. Finally, we multiply the attention mask by the decoder feature map to produce the output of the attention block.

For an image of the galaxy-quasar system, we identify five classes of components: background, central lens galaxy, quasar image, lensed arc from the quasar host galaxy, and satellite of the central lens galaxy. Therefore, the output layer for the network trained with galaxy-quasar systems has five channels, one containing the probability for each class. For an image of the galaxy-galaxy system, we simply exclude the quasar image class, thus the output layer has four channels. We apply the softmax activation function \citep{Bridle89} before the final output layer so that the values in the output channel represent the probability for the corresponding class at each pixel.

The loss function we used combines dice loss \citep{Sudre17} and focal loss \citep{Lin2018} as it is suitable for multi-class classification and imbalanced datasets. For the focal loss, we set \inlinepy{alpha = 0.25} and \inlinepy{gamma = 2} for optimal results. 

\subsection{Training dataset} \label{sec:training_set}

We create a synthetic dataset using \textsc{lenstronomy} for training. Our simulated images mimic the image quality and resolution of the \textit{Hubble} Space Telescope's (HST) Wide-Field Camera 3 (WFC3) in the F814W filter. Future upgrades to \textsc{dolphin} will deliver new NN models trained with multiple HST filters and also with ground-based imaging qualities, such as that of the Rubin Observatory LSST. We also provide our code for dataset creation and training to enable users to re-generate their own training set for the desired imaging specifications (e.g., choice of bands, ground-based or space-based resolution) and sample-level characteristics.

We use an EPL mass profile to simulate the central deflector galaxy and add an external shear field on top of it. We also add satellite deflectors with SIE mass profiles. The probability for the number of satellites in each lens system is $p(N_{\rm sat}) \propto 1/(1+N_{\rm sat})$ with $N_{\rm sat}^{\rm max} = 3$. We simulate the central and satellite deflectors' light profiles with S\'ersic functions. We simulate the source galaxy's light distribution using real spiral galaxy images from the HST legacy imaging from \citet{Shajib22}. The quasar images are simulated as point sources convolved with a realistic simulated PSF from \textsc{Tiny Tim} \citep{Krist11}. The Einstein radius for the main lens galaxy is randomly drawn from a uniform distribution between 0\farcs75 and 1\farcs5, typical for galaxy-scale lenses currently known. Similarly, the probability distributions for all the other parameters in the simulation are tuned to reproduce those observed in the galaxy-scale lens samples presented by \citet{Shajib19} and \citet{Tan24}. Finally, we add noise to the images that accounts for the background, read, and Poisson noise corresponding to 1429 s of exposure. We set the pixel scale to 0\farcs05 with the image dimension being 128$\times$128 pixels.

\begin{figure*}
    \centering
    \includegraphics[width=0.97\textwidth]{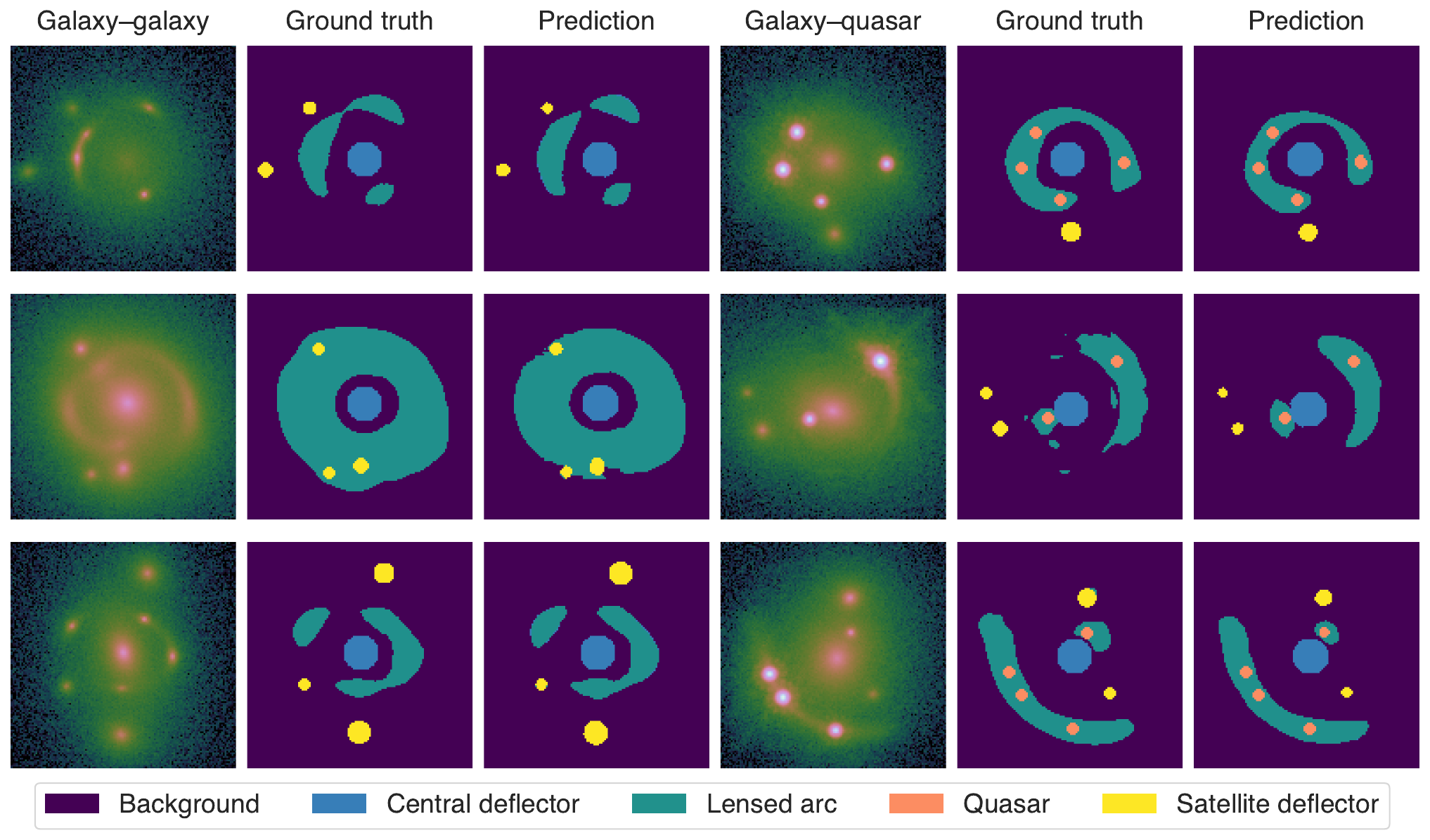}
    \caption{\label{fig:validation_test_galaxy_quasar}
        Validation results for both galaxy--galaxy (first three columns) and galaxy--quasar (last three columns) lens systems. 
        The first and fourth columns show the inputs from the validation set of synthetic images, the second and fifth columns show the ground-truth segmentation masks (i.e., labels), and the third and sixth columns show the NN-predicted segmentation masks. The color for each class in the segmentation mask is labeled in the legend. The NN model shows strong performance in segmenting the lens components, with the F1 score for each class being above 86\%.
    }
\end{figure*}

We create the segmentation mask, that is, the training labels, from the noiseless simulated images. We define five classes for the mask: central deflector, quasar host, quasar, satellite, and background. We create the labels through ad-hoc conditions applied to each class's relative flux levels and signal-to-noise ratios. Figure \ref{fig:validation_test_galaxy_quasar} illustrates some examples of the created segmentation masks (the `Ground truth' columns). The specific algorithm for this can be found in the \texttt{ai\_training/dataset.py} module of the released code suite on GitHub.




\subsection{Training procedure} \label{sec:training}

We simulated 110,000 images for each of the galaxy--galaxy and galaxy--quasar lens types. We used 90,000 of them for training, 10,000 for validation, and 10,000 for testing. We randomly remove the deflector's light from a fraction of the images (10\%) so that the network does not learn to label the central pixels within an image as the central deflector by default, given the predominance of this correlation in the training dataset. We further augment our data by applying a zoom effect up to 50\%, consistently on both the input images and labels, to generalize our model's performance for real input images that would be resized to 128$\times$128 pixels from an arbitrary size.


We train the model on the training set with the adaptive momentum \citep[Adam;][]{Kingma14} optimizer over 50 epochs, with a batch size of 32. We used early stopping with a patience of ten epochs to prevent overfitting and restore the best weights by monitoring validation loss. Additionally, we applied a learning rate reduction technique with a factor of 0.5 when the validation loss did not improve for five consecutive epochs, with a minimum learning rate of $1 \times 10^{-6}$. The model was trained using an Nvidia L4 GPU on the Google Colaboratory platform.

\subsection{Validations and tests} \label{sec:validation}

We validate the model on the test set to evaluate its performance. We calculate the precision, recall, F1 score, and Intersection over Union (IoU) for each class. The precision is the ratio of true positives to the sum of true and false positives. The recall is the ratio of true positives to the sum of true positives and false negatives. The F1 score is the harmonic mean of precision and recall. The IoU is the overlap between the predicted bounding region and the ground truth bounding region relative to their combined area. The IoU is a commonly used metric in computer vision to measure object detection accuracy. Table \ref{table:validation_metrics_quasar} shows these performance metrics for our trained NN model. The model performs strongly in segmenting the lens components, with the F1 score for each class being above 86\% 

\begin{figure*}
    \centering
    \includegraphics[width=0.97\textwidth]{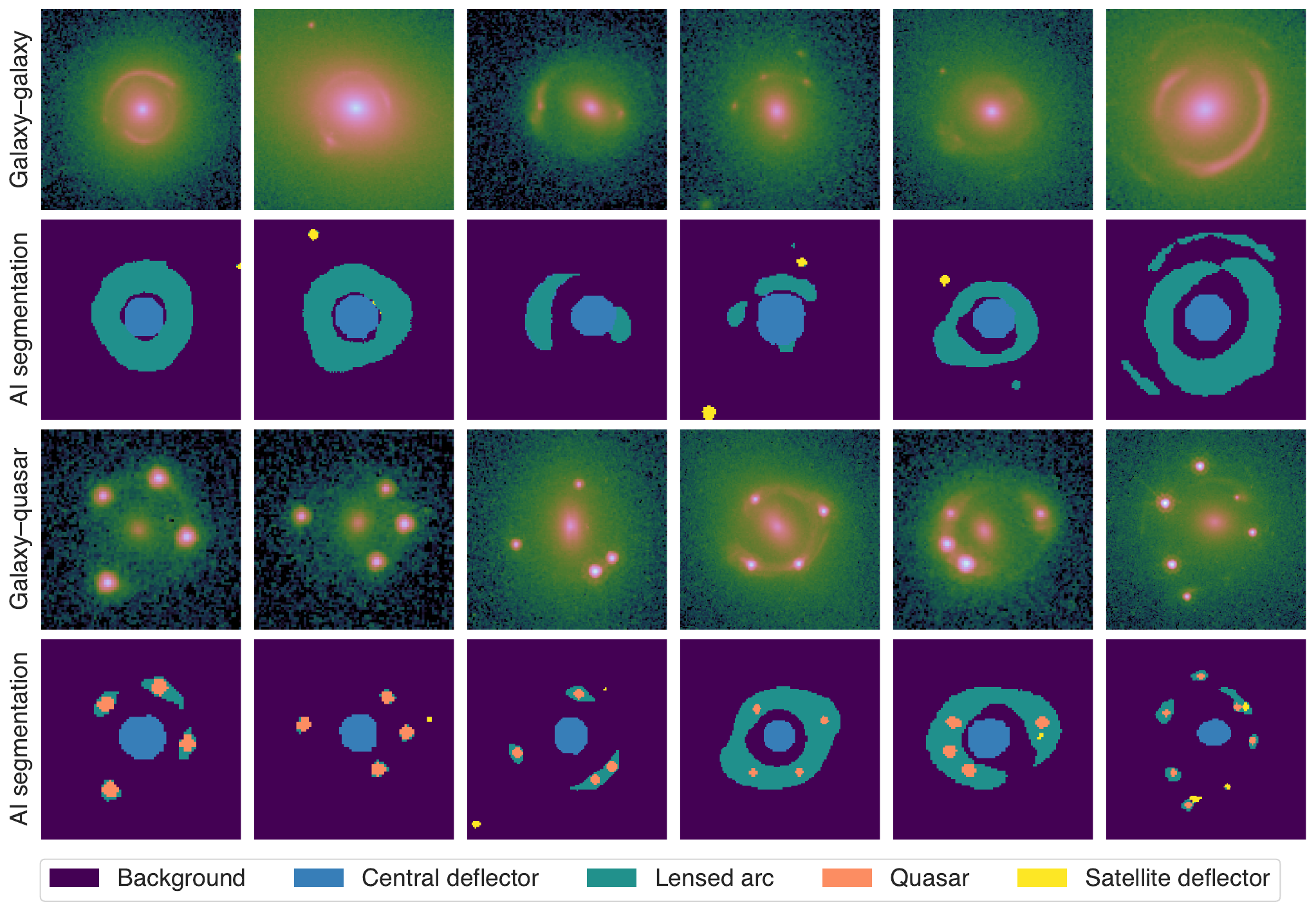}
    \caption{\label{fig:real_qso_lens_segmentation}
    Segmentation results by our trained NN model for six galaxy--galaxy lenses (top two rows) from the SLACS sample and six galaxy--quasar lenses (bottom two rows) from the STRIDES sample. Here, the input images are HST imaging in the F814W filter, with the mean background subtracted. Our NN model showcases superb performance in identifying the lens components that are labeled in the legend. The last galaxy--galaxy lens illustrated here is a compound lens, the `Jackpot' lens; and the last galaxy--quasar lens is another compound lens, the first Einstein zigzag with six images appearing of the same background quasar. Although these two highly complex systems fall outside the typical use cases for \textsc{dolphin}, as extracting their scientific richness would require well-tailored, hand-crafted models, they serve as illustrative examples of the exceptional performance of \textsc{dolphin}'s NN model in segmenting lens imaging data.
    }
\end{figure*}

In Figure \ref{fig:real_qso_lens_segmentation}, we show the segmentation results for real data for six galaxy--galaxy lenses from the Sloan Lens ACS \citep[SLACS;][]{Bolton06} sample and six galaxy--quasar lenses from the STRong lensing Insights into the Dark Energy Survey \cite[STRIDES;][]{Shajib19, Schmidt22} sample. We reshape the real HST images from their intrinsic dimension to 128$\times$128 pixels to feed them into the model. Figure \ref{fig:real_qso_lens_segmentation} illustrates the superb performance of the NN model in the semantic segmentation of the real images, where all satellite galaxies and quasar images are correctly identified in these images. For an interesting test, we provided the `Jackpot' lens \citep{Gavazzi08}, a compound lens (top row last column in Figure \ref{fig:real_qso_lens_segmentation}), and the first Einstein zigzag \citep{Dux24}, another compound lens with six images of the same background quasar. Our NN model detects both sets of arcs from the two background galaxies in the `Jackpot' lens. For the Einstein zigzag lens, our NN model correctly detects all six quasar images, but it confuses the small arcs from the intervening source galaxy as satellite deflectors. We note that the intervening source galaxy, in fact, acts as a deflector for the background quasar, which is located farther behind. However, such confusion by the NN model is not a concern here, as such complicated lenses are not expected to be modeled with an automated pipeline given the extreme level of complexity in the lensing configuration requiring carefully hand-crafted models enabling their science cases in precision cosmology (e.g., Schmidt et al., in preparation). Although these very rare \citep[1 in 500 lenses;][]{Dux24} and complicated systems will be out of the intended use case of \textsc{dolphin}, they make for illustrative examples of the superb segmentation performance of our NN model.

\begin{deluxetable*}{llccccc}
\tablewidth{0pt}
\tablecaption{\label{table:validation_metrics_quasar}
    Performance metrics of our trained NN model in semantic segmentation}
\tablehead{
\colhead{Lens type} & \colhead{Classes or components} & \colhead{Precision (\%)} & \colhead{Recall (\%)} & \colhead{F1 score (\%)} & \colhead{IoU (\%)} }
\startdata
        \hline
        Galaxy--quasar & Background & 98.42 & 99.67 & 99.04 & 98.10 \\
         & Central deflector & 91.52 & 98.67 & 94.96 & 90.41 \\
        & Lensed arc & 97.06 & 81.05 & 88.34 & 79.11 \\
        & Quasar & 97.27 & 97.51 & 97.39 & 94.91 \\
        & Satellite deflector & 91.68 & 81.90 & 86.51 & 76.23 \\
        \hline
Galaxy--galaxy         & Background      & 99.25  & 99.60 & 99.43  & 98.86  \\
& Central deflector            & 79.56  & 98.39  & 87.98  & 78.54  \\
& Lensed arc       & 97.68  & 90.01  & 93.69  & 88.13  \\
& Satellite deflector & 91.44  & 88.66  & 90.02  & 81.85  \\
\enddata
\end{deluxetable*}

\section{Discussion and conclusion} \label{sec:conclusion}

In this paper, we present \textsc{dolphin}, the first \textit{fully} automated forward modeling pipeline for galaxy-scale strong lenses, capable of modeling both galaxy--galaxy and galaxy--quasar types. \textsc{dolphin} combines an AI trained with deep learning with a traditional forward modeling engine, namely \textsc{lenstronomy}, to provide a hybrid approach to achieve full automation in lens modeling. \textsc{dolphin} performs semantic segmentation for the strong lens components with an NN model, sets up an optimal lens model using the segmentation, and then fits the model with the forward modeling engine \textsc{lenstronomy}. We provided an overview of the workflow of \textsc{dolphin} and the associated modules and a demonstration of the automated lens modeling in Section \ref{sec:overall_workflow}. We then described the NN model for visual recognition in Section \ref{sec:ml}. Our trained NN model shows strong performance in segmenting the lens components. We also showed that the model can identify the lens components in the real data images of both galaxy--galaxy and galaxy--quasar systems. This performance level in semantic segmentation of the imaging data is sufficient to set up an optimally tuned lens model for most galaxy-scale lens systems, including those that contain up to two or three satellite galaxies of the central deflector. Given the ease of use in setting up and running \textsc{dolphin}, both in personal computers and on high-performance computing clusters, \textsc{dolphin} will vastly reduce the amount of human investigator time in modeling large samples of lenses, as already demonstrated with its semi-automated (i.e., using human visual recognition) modeling mode \citep{Shajib21, Tan24, Hogg25}. Thus, the full automation achieved through the NN model presented in this paper will further reduce the human investigator time spent modeling large samples.

Like any software program, \textsc{dolphin} will receive continuous and future upgrades based on necessity and user feedback. For example, although the \texttt{Processor} class can currently perform modeling with multi-band imaging data as demonstrated in \citet{Tan24}, the NN model is currently trained only with single-band specification in the optical band (i.e., the WFC3/F814W filter). The optical band tends to be the most useful for lensing information as it sits in the sweet spot for maximizing the signal-to-noise for both the lensing and source galaxy with less blending between the two, given the `red' and `blue' colors of the lens and source galaxies, respectively. As a result, the optical (F814W) filter also tends to be the primary choice for full ML-based extraction of the lensing parameters \citep[e.g.,][]{Erickson24}. Future improvements in \textsc{dolphin}'s computer vision capability will allow multi-band imaging input to harness the color information for better detection of the lens components. In addition, a separate purpose-built network for ground-based imaging specifications will extend the pipeline's scope of usage. Furthermore, more classes of objects can be included in the training to improve the robustness of component detection, for example, stars, cosmic rays, etc., that are sometimes present in the image cutout of real systems.

The \textsc{dolphin} pipeline is especially useful for the large lens samples that will be discovered by the upcoming large area sky surveys, such as \textit{Euclid}, Rubin, and \textit{Roman} observatories. Several previous studies explored using ML algorithms for fully automated lens modeling \citep[e.g.,][]{Hezaveh17, Poh22, Erickson24}. An AI-powered forward modeler like \textsc{dolphin} has important complementarity with these ML-based algorithms and also with traditional hand-crafted lens models \citep[e.g.,][]{Birrer19b, Shajib20}. This is because these three approaches will shine in different regimes of sample sizes that are largely determined by the somewhat correlated rarity and complexity of the lens systems, the availability of high-resolution imaging, and the specific science cases in which they will be employed. More complex lens systems, for example, those with multiple lens and source planes \citep[e.g.,][]{Shajib20, Dux24}, albeit requiring hand-crafted models, are richer in information and thus their scientific usefulness justifies the manual and time-consuming modeling approach by a human investigator. A major fraction of the lensing systems have sufficiently simple lens configurations that can be modeled with \textsc{dolphin}. For example, \citet{Tan24} achieved an $\sim$80\% success rate in modeling galaxy--galaxy lenses from a considered sample size of $\mathcal{O}(10^2)$. \textsc{dolphin} is especially suitable for such sample sizes of $\mathcal{O}(10^2)$--$\mathcal{O}(10^3)$. As a forward modeling pipeline, \textsc{dolphin} will still require running sampling methods, such as the MCMC, to obtain the best-fit lens model parameters and their uncertainties. As a result, sample sizes much larger than $\mathcal{O}(10^3)$ can potentially be limited by computational resources for CPU-based modeling engines. However, GPU-accelerated, auto-differentiable modeling engines, such as \textsc{GIGA-Lens} \citep{Gu22}, \textsc{herculens} \citep{Galan22}, and \textsc{jaxtronomy}\footnote{\url{https://github.com/lenstronomy/jaxtronomy}}, will likely be up to the task. Especially, \textsc{jaxtronomy} is a direct port of \textsc{lenstronomy}, thus the modeling engine for \textsc{dolphin} can be seamlessly swapped from \textsc{lenstronomy} to \textsc{jaxtronomy}. 

In addition, ML-based algorithms will be particularly useful when it comes to sample sizes of $\mathcal{O}(10^5)$ that are forecasted to be discovered from \textit{Euclid}, Rubin, and \textit{Roman} Observatories \citep{Collett15, Shajib24b}. However, automated forward modeling pipelines like \textsc{dolphin} will be paramount to validate the ML-based models for a smaller subsample, as done by \citet{Erickson24}. In summary, we envision that all three approaches for lens modeling will be essential and complementary in the near future, with each sining at different regimes of sample sizes -- $\mathcal{O}(10)$: hand-crafted modeling for complex and scientifically rich systems, $\mathcal{O}(10^2)$--$\mathcal{O}(10^4)$: automated forward modeling with \textsc{dolphin}, and $\mathcal{O}(10^4)$--$\mathcal{O}(10^5)$: ML-based lensing parameter extraction.

\begin{acknowledgements}
    Support for this work was provided by NASA through the NASA Hubble Fellowship grant HST-HF2-51492 awarded to AJS by the Space Telescope Science Institute (STScI), which is operated by the Association of Universities for Research in Astronomy, Inc., for NASA, under contract NAS5-26555. AJS also received support from NASA through the STScI grants HST-GO-15320 and HST-GO-16773. This project was supported by the U.S. Department of Energy (DOE) Office of Science Distinguished Scientist Fellow Program. CYT was supported by NASA through the STScI grant HST-AR-16149. TT acknowledges support from the National Science Foundation through grant AST-2407277, by the Gordon and Betty Moore Foundation through grant 8548.
\end{acknowledgements}

\begin{contribution}
AJS: Conceptualization, Data curation, Funding acquisition, Project administration, Software, Visualization, Writing – original draft, Writing – review \& editing. NSN: Methodology, Writing – original draft, Writing – review \& editing. CYT: Software, Writing – review \& editing. VS: Data curation. SB: Supervision, Writing – review \& editing. TT: Supervision, Funding acquisition, Writing – review \& editing. JF: Supervision, Funding acquisition, Writing – review \& editing.
\end{contribution}

\facilities{Google Colaboratory}

 \software{
 \textsc{lenstronomy} \citep{Birrer18, Birrer21b}, \textsc{tensorflow} \citep{Abadi16}, \textsc{keras} \citep{Chollet15}, \textsc{numpy} \citep{Oliphant15}, \textsc{matplotlib} \citep{Hunter07}, \textsc{h5py} \citep{Collette13}, \textsc{seaborn} \citep{Waskom14}, \textsc{Jupyter} \citep{Kluyver16}, \textsc{PlotNeuralNetwork} \citep{Iqbal18}, \textsc{Tiny Tim} \citep{Krist11}
 }

\bibliography{ajshajib}{}

\begin{thebibliography}{}
\expandafter\ifx\csname natexlab\endcsname\relax\def\natexlab#1{#1}\fi
\providecommand{\url}[1]{\href{#1}{#1}}
\providecommand{\dodoi}[1]{doi:~\href{http://doi.org/#1}{\nolinkurl{#1}}}
\providecommand{\doeprint}[1]{\href{http://ascl.net/#1}{\nolinkurl{http://ascl.net/#1}}}
\providecommand{\doarXiv}[1]{\href{https://arxiv.org/abs/#1}{\nolinkurl{https://arxiv.org/abs/#1}}}

\bibitem[{M. Abadi {et~al.}(2016)Abadi, Barham, Chen, Chen, Davis, Dean, Devin, Ghemawat, Irving, Isard, Kudlur, Levenberg, Monga, Moore, Murray, Steiner, Tucker, Vasudevan, Warden, Wicke, Yu, \& Zheng}]{Abadi16}
Abadi, M., Barham, P., Chen, J., {et~al.} 2016, \bibinfo{title}{TensorFlow: A system for large-scale machine learning,} \dodoi{10.48550/arXiv.1605.08695}

\bibitem[{K.~T. {Abe} {et~al.}(2025){Abe}, {Oguri}, {Birrer}, {Khadka}, {Marshall}, {Lemon}, {More}, \& {LSST Dark Energy Science Collaboration}}]{Abe24}
{Abe}, K.~T., {Oguri}, M., {Birrer}, S., {et~al.} 2025, \bibinfo{title}{{A halo model approach for mock catalogs of time-variable strong gravitational lenses},} \dodoi{10.33232/001c.128482}

\bibitem[{S. Birrer \& A. Amara(2018)Birrer \& Amara}]{Birrer18}
Birrer, S., \& Amara, A. 2018, \bibinfo{title}{lenstronomy: Multi-purpose gravitational lens modelling software package,} Physics of the Dark Universe, 22, 189, \dodoi{10.1016/j.dark.2018.11.002}

\bibitem[{S. Birrer {et~al.}(2019)Birrer, Treu, Rusu, Bonvin, Fassnacht, Chan, Agnello, Shajib, Chen, Auger, Courbin, Hilbert, Sluse, Suyu, Wong, Marshall, Lemaux, \& Meylan}]{Birrer19b}
Birrer, S., Treu, T., Rusu, C.~E., {et~al.} 2019, \bibinfo{title}{H0LiCOW - IX. Cosmographic analysis of the doubly imaged quasar SDSS 1206+4332 and a new measurement of the Hubble constant,} \mnras, 484, 4726, \dodoi{10.1093/mnras/stz200}

\bibitem[{S. Birrer {et~al.}(2021)Birrer, Shajib, Gilman, Galan, Aalbers, Millon, Morgan, Pagano, Park, Teodori, Tessore, Ueland, Vyvere, Wagner-Carena, Wempe, Yang, Ding, Schmidt, Sluse, Zhang, \& Amara}]{Birrer21b}
Birrer, S., Shajib, A.~J., Gilman, D., {et~al.} 2021, \bibinfo{title}{lenstronomy II: A gravitational lensing software ecosystem,} JOSS, 6, 3283, \dodoi{10.21105/joss.03283}

\bibitem[{S. Birrer {et~al.}(2024)Birrer, Millon, Sluse, Shajib, Courbin, Erickson, Koopmans, Suyu, \& Treu}]{Birrer24}
Birrer, S., Millon, M., Sluse, D., {et~al.} 2024, \bibinfo{title}{Time-Delay Cosmography: Measuring the Hubble Constant and Other Cosmological Parameters with Strong Gravitational Lensing,} Space Science Reviews, 220, 48, \dodoi{10.1007/s11214-024-01079-w}

\bibitem[{A.~S. Bolton {et~al.}(2006)Bolton, Burles, Koopmans, Treu, \& Moustakas}]{Bolton06}
Bolton, A.~S., Burles, S., Koopmans, L. V.~E., Treu, T., \& Moustakas, L.~A. 2006, \bibinfo{title}{The Sloan Lens ACS Survey. I. A Large Spectroscopically Selected Sample of Massive Early-Type Lens Galaxies,} \apj, 638, 703, \dodoi{10.1086/498884}

\bibitem[{J. Bridle(1989)Bridle}]{Bridle89}
Bridle, J. 1989, in Advances in Neural Information Processing Systems, Vol.~2 (Morgan-Kaufmann).
\newblock \url{https://proceedings.neurips.cc/paper/1989/hash/0336dcbab05b9d5ad24f4333c7658a0e-Abstract.html}

\bibitem[{F. Chollet {et~al.}(2015)Chollet {et~al.}}]{Chollet15}
Chollet, F., {et~al.} 2015, \bibinfo{title}{Keras,} \url{https://keras.io}

\bibitem[{T.~E. Collett(2015)Collett}]{Collett15}
Collett, T.~E. 2015, \bibinfo{title}{The Population of Galaxy-Galaxy Strong Lenses in Forthcoming Optical Imaging Surveys,} \apj, 811, 20, \dodoi{10.1088/0004-637X/811/1/20}

\bibitem[{A. Collette(2013)Collette}]{Collette13}
Collette, A. 2013, Python and HDF5 (O'Reilly)

\bibitem[{F. {Dux} {et~al.}(2025){Dux}, {Millon}, {Lemon}, {Schmidt}, {Courbin}, {Shajib}, {Treu}, {Birrer}, {Wong}, {Agnello}, {Andrade}, {Galan}, {Hjorth}, {Paic}, {Schuldt}, {Schweinfurth}, {Sluse}, {Smette}, \& {Suyu}}]{Dux24}
{Dux}, F., {Millon}, M., {Lemon}, C., {et~al.} 2025, \bibinfo{title}{{J1721+8842: The first Einstein zigzag lens},} \aap, 694, A300, \dodoi{10.1051/0004-6361/202452970}

\bibitem[{S. Erickson {et~al.}(2024)Erickson, Wagner-Carena, Marshall, Millon, Birrer, Roodman, Schmidt, Treu, Schuldt, Shajib, Venkatraman, \& Collaboration}]{Erickson24}
Erickson, S., Wagner-Carena, S., Marshall, P., {et~al.} 2024, \bibinfo{title}{Lens Modeling of STRIDES Strongly Lensed Quasars using Neural Posterior Estimation,} \dodoi{10.48550/arXiv.2410.10123}

\bibitem[{S. {Ertl} {et~al.}(2023){Ertl}, {Schuldt}, {Suyu}, {Schmidt}, {Treu}, {Birrer}, {Shajib}, \& {Sluse}}]{Ertl23}
{Ertl}, S., {Schuldt}, S., {Suyu}, S.~H., {et~al.} 2023, \bibinfo{title}{{TDCOSMO. X. Automated modeling of nine strongly lensed quasars and comparison between lens-modeling software},} \aap, 672, A2, \dodoi{10.1051/0004-6361/202244909}

\bibitem[{ {Euclid Collaboration} {et~al.}(2025){Euclid Collaboration}, {Busillo}, {Tortora}, {Metcalf}, {Nightingale}, {Meneghetti}, {Gentile}, {Gavazzi}, {Zhong}, {Li}, {Cl{\'e}ment}, {Covone}, {Napolitano}, {Courbin}, {Walmsley}, {Jullo}, {Pearson}, {Scott}, {Le Brun}, {Leuzzi}, {Aghanim}, {Altieri}, {Amara}, {Andreon}, {Aussel}, {Baccigalupi}, {Baldi}, {Bardelli}, {Battaglia}, {Biviano}, {Branchini}, {Brescia}, {Brinchmann}, {Camera}, {Ca{\~n}as-Herrera}, {Capobianco}, {Carbone}, {Cardone}, {Carretero}, {Casas}, {Castellano}, {Castignani}, {Cavuoti}, {Chambers}, {Cimatti}, {Colodro-Conde}, {Congedo}, {Conselice}, {Conversi}, {Copin}, {Courtois}, {Cropper}, {Da Silva}, {Degaudenzi}, {de la Torre}, {De Lucia}, {Di Giorgio}, {Dinis}, {Dole}, {Dubath}, {Dupac}, {Dusini}, {Escoffier}, {Farina}, {Farinelli}, {Faustini}, {Ferriol}, {Finelli}, {Fotopoulou}, {Frailis}, {Franceschi}, {Galeotta}, {George}, {Gillard}, {Gillis}, {Giocoli}, {Gracia-Carpio}, {Granett}, {Grazian}, {Grupp}, {Haugan}, {Holmes}, {Hook}, {Hormuth}, {Hornstrup}, {Hudelot}, {Jahnke}, {Jhabvala}, {Joachimi}, {Keih{\"a}nen}, {Kermiche}, {Kiessling}, {Kubik}, {K{\"u}mmel}, {Kunz}, {Kurki-Suonio}, {Le Boulc'h}, {Ligori}, {Lilje}, {Lindholm}, {Lloro}, {Mainetti}, {Maino}, {Maiorano}, {Mansutti}, {Marggraf}, {Markovic}, {Martinelli}, {Martinet}, {Marulli}, {Massey}, {Maurogordato}, {Medinaceli}, {Mei}, {Mellier}, {Merlin}, {Meylan}, {Mora}, {Moresco}, {Moscardini}, {Nakajima}, {Neissner}, {Niemi}, {Padilla}, {Paltani}, {Pasian}, {Pedersen}, {Pettorino}, {Pires}, {Polenta}, {Poncet}, {Popa}, {Pozzetti}, {Raison}, {Rebolo}, {Renzi}, {Rhodes}, {Riccio}, {Romelli}, {Roncarelli}, {Saglia}, {Sakr}, {S{\'a}nchez}, {Sapone}, {Sartoris}, {Schewtschenko}, {Schirmer}, {Schneider}, {Schrabback}, {Secroun}, {Sefusatti}, {Seidel}, {Seiffert}, {Serrano}, {Simon}, {Sirignano}, {Sirri}, {Smadja}, {Stanco}, {Steinwagner}, {Tallada-Cresp{\'\i}}, {Taylor}, {Tereno}, {Toft}, {Toledo-Moreo}, {Torradeflot}, {Tutusaus}, {Valenziano}, {Valiviita}, {Vassallo}, {Veropalumbo},
  {Wang}, {Weller}, {Zamorani}, {Zucca}, {Allevato}, {Ballardini}, {Bolzonella}, {Bozzo}, {Burigana}, {Cabanac}, {Calabrese}, {Di Ferdinando}, {Escartin Vigo}, {Gabarra}, {Huertas-Company}, {Matthew}, {Mauri}, {Nucita}, {Pezzotta}, {P{\"o}ntinen}, {Porciani}, {Scottez}, {Tenti}, {Viel}, {Wiesmann}, {Akrami}, {Alvi}, {Andika}, \& {Anselmi}}]{Euclid25}
{Euclid Collaboration}, {Busillo}, V., {Tortora}, C., {et~al.} 2025, \bibinfo{title}{{Euclid Quick Data Release (Q1). LEMON -- Lens Modelling with Neural networks. Automated and fast modelling of Euclid gravitational lenses with a singular isothermal ellipsoid mass profile},} arXiv e-prints, arXiv:2503.15329, \dodoi{10.48550/arXiv.2503.15329}

\bibitem[{A. {Galan} {et~al.}(2022){Galan}, {Vernardos}, {Peel}, {Courbin}, \& {Starck}}]{Galan22}
{Galan}, A., {Vernardos}, G., {Peel}, A., {Courbin}, F., \& {Starck}, J.~L. 2022, \bibinfo{title}{{Using wavelets to capture deviations from smoothness in galaxy-scale strong lenses},} \aap, 668, A155, \dodoi{10.1051/0004-6361/202244464}

\bibitem[{R. Gavazzi {et~al.}(2008)Gavazzi, Treu, Koopmans, Bolton, Moustakas, Burles, \& Marshall}]{Gavazzi08}
Gavazzi, R., Treu, T., Koopmans, L. V.~E., {et~al.} 2008, \bibinfo{title}{The Sloan Lens ACS Survey. VI. Discovery and Analysis of a Double Einstein Ring,} \apj, 677, 1046, \dodoi{10.1086/529541}

\bibitem[{A. {Gu} {et~al.}(2022){Gu}, {Huang}, {Sheu}, {Aldering}, {Bolton}, {Boone}, {Dey}, {Filipp}, {Jullo}, {Perlmutter}, {Rubin}, {Schlafly}, {Schlegel}, {Shu}, \& {Suyu}}]{Gu22}
{Gu}, A., {Huang}, X., {Sheu}, W., {et~al.} 2022, \bibinfo{title}{{GIGA-Lens: Fast Bayesian Inference for Strong Gravitational Lens Modeling},} \apj, 935, 49, \dodoi{10.3847/1538-4357/ac6de4}

\bibitem[{Y.~D. Hezaveh {et~al.}(2017)Hezaveh, Levasseur, \& Marshall}]{Hezaveh17}
Hezaveh, Y.~D., Levasseur, L.~P., \& Marshall, P.~J. 2017, \bibinfo{title}{Fast automated analysis of strong gravitational lenses with convolutional neural networks,} \nat, 548, 555, \dodoi{10.1038/nature23463}

\bibitem[{N.~B. {Hogg} {et~al.}(2025){Hogg}, {Shajib}, {Johnson}, \& {Larena}}]{Hogg25}
{Hogg}, N.~B., {Shajib}, A.~J., {Johnson}, D., \& {Larena}, J. 2025, \bibinfo{title}{{Line-of-sight shear in SLACS strong lenses},} arXiv e-prints, arXiv:2501.16292, \dodoi{10.48550/arXiv.2501.16292}

\bibitem[{J.~D. Hunter(2007)Hunter}]{Hunter07}
Hunter, J.~D. 2007, \bibinfo{title}{Matplotlib: A 2D Graphics Environment,} Computing in Science and Engineering, 9, 90, \dodoi{10.1109/MCSE.2007.55}

\bibitem[{H. Iqbal(2018)Iqbal}]{Iqbal18}
Iqbal, H. 2018, \bibinfo{title}{HarisIqbal88/PlotNeuralNet v1.0.0,}, v1.0.0 Zenodo, \dodoi{10.5281/zenodo.2526396}

\bibitem[{J. Kennedy \& R. Eberhart(1995)Kennedy \& Eberhart}]{Kennedy95}
Kennedy, J., \& Eberhart, R. 1995, in Proceedings of ICNN\textquotesingle95 - International Conference on Neural Networks (IEEE), \dodoi{10.1109/icnn.1995.488968}

\bibitem[{D.~P. Kingma \& J. Ba(2014)Kingma \& Ba}]{Kingma14}
Kingma, D.~P., \& Ba, J. 2014, \bibinfo{title}{Adam: A Method for Stochastic Optimization,} arXiv, \dodoi{10.48550/arXiv.1412.6980}

\bibitem[{T. Kluyver {et~al.}(2016)Kluyver, Ragan-Kelley, Pérez, Granger, Bussonnier, Frederic, Kelley, Hamrick, Grout, Corlay, Ivanov, Avila, Abdalla, \& Willing}]{Kluyver16}
Kluyver, T., Ragan-Kelley, B., Pérez, F., {et~al.} 2016, in Positioning and Power in Academic Publishing: Players, Agents and Agendas, ed. F.~Loizides \& B.~Schmidt (IOS Press BV, Amsterdam, Netherlands), 87 -- 90, \dodoi{10.3233/978-1-61499-649-1-87}

\bibitem[{J.-P. Kneib {et~al.}(2011)Kneib, Bonnet, Golse, Sand, Jullo, \& Marshall}]{Kneib11}
Kneib, J.-P., Bonnet, H., Golse, G., {et~al.} 2011, \bibinfo{title}{LENSTOOL: A Gravitational Lensing Software for Modeling Mass Distribution of Galaxies and Clusters (strong and weak regime),} Astrophysics Source Code Library, ascl:1102.004

\bibitem[{R. Kormann {et~al.}(1994)Kormann, Schneider, \& Bartelmann}]{Kormann94}
Kormann, R., Schneider, P., \& Bartelmann, M. 1994, \bibinfo{title}{Isothermal elliptical gravitational lens models,} \aap, 284, 285

\bibitem[{J.~E. Krist {et~al.}(2011)Krist, Hook, \& Stoehr}]{Krist11}
Krist, J.~E., Hook, R.~N., \& Stoehr, F. 2011, in Society of Photo-Optical Instrumentation Engineers (SPIE) Conference Series, Vol. 8127, \procspie, 81270J, \dodoi{10.1117/12.892762}

\bibitem[{C. {Lemon} {et~al.}(2023){Lemon}, {Anguita}, {Auger-Williams}, {Courbin}, {Galan}, {McMahon}, {Neira}, {Oguri}, {Schechter}, {Shajib}, {Treu}, {Agnello}, \& {Spiniello}}]{Lemon23}
{Lemon}, C., {Anguita}, T., {Auger-Williams}, M.~W., {et~al.} 2023, \bibinfo{title}{{Gravitationally lensed quasars in Gaia - IV. 150 new lenses, quasar pairs, and projected quasars},} \mnras, 520, 3305, \dodoi{10.1093/mnras/stac3721}

\bibitem[{T.-Y. Lin {et~al.}(2018)Lin, Goyal, Girshick, He, \& Dollár}]{Lin2018}
Lin, T.-Y., Goyal, P., Girshick, R., He, K., \& Dollár, P. 2018, \bibinfo{title}{Focal Loss for Dense Object Detection,} \doarXiv{1708.02002}

\bibitem[{V. Nair \& G.~E. Hinton(2010)Nair \& Hinton}]{Nair10}
Nair, V., \& Hinton, G.~E. 2010, in Proceedings of the 27th International Conference on International Conference on Machine Learning, ICML'10 (Madison, WI, USA: Omnipress), 807–814

\bibitem[{J. Nightingale {et~al.}(2018)Nightingale, Dye, \& Massey}]{Nightingale18}
Nightingale, J., Dye, S., \& Massey, R. 2018, \bibinfo{title}{AutoLens: automated modeling of a strong lens's light, mass, and source,} \mnras, 478, 4738, \dodoi{10.1093/mnras/sty1264}

\bibitem[{M. Oguri(2010)Oguri}]{Oguri10b}
Oguri, M. 2010, \bibinfo{title}{The Mass Distribution of SDSS J1004+4112 Revisited,} Publications of the Astronomical Society of Japan, 62, 1017, \dodoi{10.1093/pasj/62.4.1017}

\bibitem[{M. Oguri \& P.~J. Marshall(2010)Oguri \& Marshall}]{Oguri10}
Oguri, M., \& Marshall, P.~J. 2010, \bibinfo{title}{Gravitationally lensed quasars and supernovae in future wide-field optical imaging surveys,} \mnras, 405, 2579, \dodoi{10.1111/j.1365-2966.2010.16639.x}

\bibitem[{O. Oktay {et~al.}(2018)Oktay, Schlemper, Le~Folgoc, Lee, Heinrich, Misawa, Mori, McDonagh, Hammerla, Kainz, Glocker, \& Rueckert}]{Oktay2018}
Oktay, O., Schlemper, J., Le~Folgoc, L., {et~al.} 2018, \bibinfo{title}{Attention U-Net: Learning Where to Look for the Pancreas,} \doarXiv{1804.03999}

\bibitem[{T.~E. Oliphant(2015)Oliphant}]{Oliphant15}
Oliphant, T.~E. 2015, Guide to NumPy, 2nd edn. (USA: CreateSpace Independent Publishing Platform)

\bibitem[{J. Poh {et~al.}(2022)Poh, Samudre, Ćiprijanović, Nord, Khullar, Tanoglidis, \& Frieman}]{Poh22}
Poh, J., Samudre, A., Ćiprijanović, A., {et~al.} 2022, \bibinfo{title}{Strong Lensing Parameter Estimation on Ground-Based Imaging Data Using Simulation-Based Inference,} \dodoi{10.48550/arXiv.2211.05836}

\bibitem[{O. {Ronneberger} {et~al.}(2015){Ronneberger}, {Fischer}, \& {Brox}}]{Ronneberger15}
{Ronneberger}, O., {Fischer}, P., \& {Brox}, T. 2015, \bibinfo{title}{{U-Net: Convolutional Networks for Biomedical Image Segmentation},} arXiv e-prints, arXiv:1505.04597, \dodoi{10.48550/arXiv.1505.04597}

\bibitem[{T. {Schmidt} {et~al.}(2023){Schmidt}, {Treu}, {Birrer}, {Shajib}, {Lemon}, {Millon}, {Sluse}, {Agnello}, {Anguita}, {Auger-Williams}, {McMahon}, {Motta}, {Schechter}, {Spiniello}, {Kayo}, {Courbin}, {Ertl}, {Fassnacht}, {Frieman}, {More}, {Schuldt}, {Suyu}, {Aguena}, {Andrade-Oliveira}, {Annis}, {Bacon}, {Bertin}, {Brooks}, {Burke}, {Carnero Rosell}, {Carrasco Kind}, {Carretero}, {Conselice}, {Costanzi}, {da Costa}, {Pereira}, {De Vicente}, {Desai}, {Doel}, {Everett}, {Ferrero}, {Friedel}, {Garc{\'\i}a-Bellido}, {Gaztanaga}, {Gruen}, {Gruendl}, {Gschwend}, {Gutierrez}, {Hinton}, {Hollowood}, {Honscheid}, {James}, {Kuehn}, {Lahav}, {Menanteau}, {Miquel}, {Palmese}, {Paz-Chinch{\'o}n}, {Pieres}, {Plazas Malag{\'o}n}, {Prat}, {Rodriguez-Monroy}, {Romer}, {Sanchez}, {Scarpine}, {Sevilla-Noarbe}, {Smith}, {Suchyta}, {Tarle}, {To}, {Varga}, \& {DES Collaboration}}]{Schmidt22}
{Schmidt}, T., {Treu}, T., {Birrer}, S., {et~al.} 2023, \bibinfo{title}{{STRIDES: automated uniform models for 30 quadruply imaged quasars},} \mnras, 518, 1260, \dodoi{10.1093/mnras/stac2235}

\bibitem[{A.~J. Shajib {et~al.}(2024{\natexlab{a}})Shajib, Smith, Birrer, Verma, Arendse, \& Collett}]{Shajib24b}
Shajib, A.~J., Smith, G.~P., Birrer, S., {et~al.} 2024{\natexlab{a}}, \bibinfo{title}{Strong gravitational lenses from the Vera C. Rubin Observatory,} \dodoi{10.48550/arXiv.2406.08919}

\bibitem[{A.~J. Shajib {et~al.}(2021)Shajib, Treu, Birrer, \& Sonnenfeld}]{Shajib21}
Shajib, A.~J., Treu, T., Birrer, S., \& Sonnenfeld, A. 2021, \bibinfo{title}{Dark matter haloes of massive elliptical galaxies at z $\sim$ 0.2 are well described by the Navarro–Frenk–White profile,} \mnras, 503, 2380, \dodoi{10.1093/mnras/stab536}

\bibitem[{A.~J. Shajib {et~al.}(2019)Shajib, Birrer, Treu, Auger, Agnello, Anguita, Buckley-Geer, Chan, Collett, Courbin, Fassnacht, Frieman, Kayo, Lemon, Lin, Marshall, McMahon, More, Morgan, Motta, Oguri, Ostrovski, Rusu, Schechter, Shanks, Suyu, Meylan, Abbott, Allam, Annis, Avila, Bertin, Brooks, Carnero~Rosell, Carrasco~Kind, Carretero, Cunha, da~Costa, De~Vicente, Desai, Doel, Flaugher, Fosalba, García-Bellido, Gerdes, Gruen, Gruendl, Gutierrez, Hartley, Hollowood, Hoyle, James, Kuehn, Kuropatkin, Lahav, Lima, Maia, March, Marshall, Melchior, Menanteau, Miquel, Plazas, Sanchez, Scarpine, Sevilla-Noarbe, Smith, Soares-Santos, Sobreira, Suchyta, Swanson, Tarle, \& Walker}]{Shajib19}
Shajib, A.~J., Birrer, S., Treu, T., {et~al.} 2019, \bibinfo{title}{Is every strong lens model unhappy in its own way? Uniform modelling of a sample of 13 quadruply+ imaged quasars,} \mnras, 483, 5649, \dodoi{10.1093/mnras/sty3397}

\bibitem[{A.~J. Shajib {et~al.}(2020)Shajib, Birrer, Treu, Agnello, Buckley-Geer, Chan, Christensen, Lemon, Lin, Millon, Poh, Rusu, Sluse, Spiniello, Chen, Collett, Courbin, Fassnacht, Frieman, Galan, Gilman, More, Anguita, Auger, Bonvin, McMahon, Meylan, Wong, Abbott, Annis, Avila, Bechtol, Brooks, Brout, Burke, Carnero~Rosell, Carrasco~Kind, Carretero, Castander, Costanzi, da~Costa, De~Vicente, Desai, Dietrich, Doel, Drlica-Wagner, Evrard, Finley, Flaugher, Fosalba, García-Bellido, Gerdes, Gruen, Gruendl, Gschwend, Gutierrez, Hollowood, Honscheid, Huterer, James, Jeltema, Krause, Kuropatkin, Li, Lima, MacCrann, Maia, Marshall, Melchior, Miquel, Ogando, Palmese, Paz-Chinchón, Plazas, Romer, Roodman, Sako, Sanchez, Santiago, Scarpine, Schubnell, Scolnic, Serrano, Sevilla-Noarbe, Smith, Soares-Santos, Suchyta, Tarle, Thomas, Walker, \& Zhang}]{Shajib20}
Shajib, A.~J., Birrer, S., Treu, T., {et~al.} 2020, \bibinfo{title}{STRIDES: a 3.9 per cent measurement of the Hubble constant from the strong lens system DES J0408-5354,} \mnras, 494, 6072, \dodoi{10.1093/mnras/staa828}

\bibitem[{A.~J. Shajib {et~al.}(2022)Shajib, Glazebrook, Barone, Lewis, Jones, Tran, Buckley-Geer, Collett, Frieman, \& Jacobs}]{Shajib22}
Shajib, A.~J., Glazebrook, K., Barone, T., {et~al.} 2022, \bibinfo{title}{LensingETC: A Tool to Optimize Multifilter Imaging Campaigns of Galaxy-scale Strong Lensing Systems,} The Astrophysical Journal, 938, 141, \dodoi{10.3847/1538-4357/ac927b}

\bibitem[{A.~J. Shajib {et~al.}(2024{\natexlab{b}})Shajib, Vernardos, Collett, Motta, Sluse, Williams, Saha, Birrer, Spiniello, \& Treu}]{Shajib24}
Shajib, A.~J., Vernardos, G., Collett, T.~E., {et~al.} 2024{\natexlab{b}}, \bibinfo{title}{Strong Lensing by Galaxies,} Space Science Reviews, 220, 87, \dodoi{10.1007/s11214-024-01105-x}

\bibitem[{C.~H. Sudre {et~al.}(2017)Sudre, Li, Vercauteren, Ourselin, \& Cardoso}]{Sudre17}
Sudre, C.~H., Li, W., Vercauteren, T., Ourselin, S., \& Cardoso, M.~J. 2017, in Deep Learning in Medical Image Analysis and Multimodal Learning for Clinical Decision Support (Springer), 240--248, \dodoi{10.1007/978-3-319-67558-9_28}

\bibitem[{S.~H. Suyu {et~al.}(2010)Suyu, Marshall, Auger, Hilbert, Blandford, Koopmans, Fassnacht, \& Treu}]{Suyu10}
Suyu, S.~H., Marshall, P.~J., Auger, M.~W., {et~al.} 2010, \bibinfo{title}{Dissecting the Gravitational lens B1608+656. II. Precision Measurements of the Hubble Constant, Spatial Curvature, and the Dark Energy Equation of State,} \apj, 711, 201, \dodoi{10.1088/0004-637X/711/1/201}

\bibitem[{J.~L. Sérsic(1968)Sérsic}]{Sersic68}
Sérsic, J.~L. 1968, Atlas de Galaxias Australes.
\newblock \url{http://adsabs.harvard.edu/abs/1968adga.book.....S}

\bibitem[{C.~Y. Tan {et~al.}(2024)Tan, Shajib, Birrer, Sonnenfeld, Treu, Wells, Williams, Buckley-Geer, Drlica-Wagner, \& Frieman}]{Tan24}
Tan, C.~Y., Shajib, A.~J., Birrer, S., {et~al.} 2024, \bibinfo{title}{Project Dinos I: A joint lensing-dynamics constraint on the deviation from the power law in the mass profile of massive ellipticals,} Monthly Notices of the Royal Astronomical Society, 530, 1474, \dodoi{10.1093/mnras/stae884}

\bibitem[{N. Tessore \& R.~B. Metcalf(2015)Tessore \& Metcalf}]{Tessore15}
Tessore, N., \& Metcalf, R.~B. 2015, \bibinfo{title}{The elliptical power law profile lens,} \aap, 580, A79, \dodoi{10.1051/0004-6361/201526773}

\bibitem[{S. Vegetti {et~al.}(2024)Vegetti, Birrer, Despali, Fassnacht, Gilman, Hezaveh, Perreault Levasseur, McKean, Powell, O’Riordan, \& Vernardos}]{Vegetti24}
Vegetti, S., Birrer, S., Despali, G., {et~al.} 2024, \bibinfo{title}{Strong Gravitational Lensing as a Probe of Dark Matter,} Space Science Reviews, 220, 58, \dodoi{10.1007/s11214-024-01087-w}

\bibitem[{M. Waskom {et~al.}(2014)Waskom, Botvinnik, Hobson, Cole, Halchenko, Hoyer, Miles, Augspurger, Yarkoni, Megies, Coelho, Wehner, cynddl, Ziegler, diego0020, Zaytsev, Hoppe, Seabold, Cloud, Koskinen, Meyer, Qalieh, \& Allan}]{Waskom14}
Waskom, M., Botvinnik, O., Hobson, P., {et~al.} 2014, seaborn: v0.5.0 (November 2014), \dodoi{10.5281/zenodo.12710}

\end{thebibliography}
\bibliographystyle{aasjournalv7}

\end{document}